\documentclass[10pt,letterpaper]{article}
\usepackage[font={footnotesize}]{caption}
\usepackage{cite}
\usepackage{graphicx}
\usepackage{tabularx}
\usepackage{bm}
\usepackage{amsmath}
\usepackage{esint}

\newcommand{ \maxm }{\mathop{\mbox{max}}}
\newcommand{ \real }{\mathop{\mbox{Re}}}
\newcommand{ \imag }{\mathop{\mbox{Im}}}
\newcommand{\bra}[1]{\langle #1|}
\newcommand{\ket}[1]{|#1\rangle}

\newcommand{\apl}{Appl. Phys. Lett.}
\newcommand{\ol}{Opt. Lett.}
\newcommand{\opex}{Opt. Express}
\newcommand{\prl}{Phys. Rev. Lett.}
\newcommand{\pra}{Phys. Rev. A}
\newcommand{\prb}{Phys. Rev. B}
\newcommand{\pre}{Phys. Rev. E}
\newcommand{\nat}{Nature}

\begin{document}

\title{Transfer of arbitrary quantum emitter states to near-field photon superpositions in nanocavities}

\author{Arthur~C.~T.~Thijssen\thanks{H.H. Wills Physics Laboratory, University of Bristol, Tyndall Avenue, Bristol BS8 1TL, UK}, Martin~J.~Cryan\thanks{Merchant Venturers School of Engineering, University of Bristol, Woodland Road, Bristol BS8 1TL, UK},
John~G.~Rarity,$^{\dagger}$ and Ruth~Oulton$^{*,\dagger}$}

\maketitle

\begin{abstract}
We present a method to analyze the suitability of particular photonic cavity designs for information exchange between arbitrary superposition states of a quantum emitter and the near-field photonic cavity mode.  As an illustrative example, we consider whether quantum dot emitters embedded in ``L3'' and ``H1'' photonic crystal cavities are able to transfer a spin superposition state to a confined photonic superposition state for use in quantum information transfer. Using an established dyadic Green's function (DGF) analysis, we describe methods to calculate coupling to arbitrary quantum emitter positions and orientations using the modified local density of states (LDOS) calculated using numerical finite-difference time-domain (FDTD) simulations.  We find that while superposition states are not supported in L3 cavities, the double degeneracy of the H1 cavities supports superposition states of the two orthogonal modes that may be described as states on a Poincar\'{e}-like sphere.  Methods are developed to comprehensively analyze the confined superposition state generated from an arbitrary emitter position and emitter dipole orientation.
\end{abstract}

%%%%%%%%%%%%%%%%%%%%%%%%%%  body  %%%%%%%%%%%%%%%%%%%%%%%%%%
\section{Introduction}
Quantum emitters such as self-assembled III-V quantum dots (QDs) have long been proposed as a key component in quantum computing architectures. Many of the specific applications of these QDs rely on transferring information between the polarization of a photon and the carrier spins in exciton states, that are made possible due to strict atomic-like selection rules.  The selection rules governing interaction of photons with the QD ground state transition are well-defined. Due to angular momentum conservation, a circularly polarized photon with right (left) helicity will only interact with an electron in the $| \downarrow \rangle$,$| \uparrow \rangle$ spin $\pm1/2$ state and a hole in the $| \Uparrow \rangle$, $| \Downarrow \rangle$ $\pm3/2$ spin state, respectively, giving angular momentum change $\Delta l = \pm 1$.

This selection rule is the basis of schemes to use QDs to store and manipulate photon polarization states using carrier spins. Of several widely-proposed components we discuss one in this paper. A QD spin-photon interface \cite{Hu:Giant_Faraday_Rotation:PRB} relies on the optical transitions between a resident electron spin and a charged exciton (or ``trion'') state. Indeed, any superposition of states on the electron spin Bloch sphere, $\alpha|\uparrow\rangle + \beta|\downarrow\rangle$ has a one-to-one correspondence with a photon polarization state on the Poincar\'{e} sphere, $\alpha|L\rangle + \beta|R\rangle$, where $\alpha$ and $\beta$ are complex, and contain the crucial phase information for quantum information.  The selection rules have been shown to be accurate for light interacting with the QD in the paraxial approximation, and in a vacuum with uniform local density of states (LDOS).

We note also that the same approach may be used for another application: a QD entangled photon pair source produced via the cascaded decay of the QD biexciton-exciton \cite{Dousse:Ultra_bright_entanglement_photon_source:Nature}. This source produces two photons entangled in linear polarization, whose fidelity is dependent on the $x$ and $y$ exciton decay paths being indistinguishable, adding an extra complication to the analysis \cite{Larque:H1_cavities_photon_pairs:NJP}.

In order to increase the light-matter interaction strength for the high demands of quantum information technology the QDs must be incorporated into a photonic structure.  Along with micropillar cavities \cite{Schneider:Site_control_rev:Nanotech}, photonic crystal cavities (PhCCs) have become a popular means to increase light-matter interaction.  Their high quality (Q)-factors and low modal volumes have already allowed the strong coupling regime to be reached \cite{Yoshie:Strong_Coupling:Nature}.

The use of these PhCCs for any applications which rely on the spin selection rules is problematic, however.  The light matter interaction is increased by strongly modifying the LDOS in the cavity. The cavity produces a highly anisotropic LDOS.  This means that a QD interacting with the LDOS couple to one polarization state more strongly than the other. Moreover, the phase information contained in the spin state $\alpha|\uparrow\rangle + \beta|\downarrow\rangle$ may be lost if the coupling to the $|\uparrow \rangle$ and $|\downarrow \rangle$ differs.

The QD interacts with the intrinsic modes of the cavity, which also have a polarization-dependent spatial variation.  The position of the QD is also crucial: the E-field spatial distribution of each polarization component differs, even for the case where the cavity produces two modes degenerate in energy, as we demonstrate.
Finally, the trapped photon then eventually either leaks into the far-field outside the structure, or is transmitted down a waveguide.  The polarization of the light collected and measured therefore depends on the difference in photon lifetime and angular emission for each polarization component.

Thus it is clear that coupling photons to spin polarization states in cavities is more complex than in the paraxial approximation, where a simple one-to-one correspondence between the Bloch and Poincar\'{e} spheres exists.  In this paper we discuss a methodology to analyze PhCC designs for their suitability for spin-dependent quantum information processes in the weak-coupling regime, taking as examples well-known ``L3'' and ``H1'' cavity designs.

We will discuss that in order to use photonic crystal cavities to transfer spin-polarization information with high fidelity, the cavity must meet two criteria. The first, Criterion 1, is that the cavity should have doubly-degenerate modes (or at least spectrally overlapping modes). The doubly-degenerate modes are orthogonal states, allowing an arbitrary superposition state to be generated. This allows one to build up a Poincar\'{e}-like sphere, such that one is able to excite the two modes (equivalent to the ``H'' and ``V'' linear polarization states in the paraxial approximation) with arbitrary intensity and phase. In fact, this criterion has been addressed experimentally by applying various methods to tune fabrication imperfections in H1 cavities to produce degenerate modes \cite{Hennessy:AFM_oxidation:APL,Luxmoore:strain_tuning:APL}.  Criterion 2 is that the modes, which have a complex structure in the x-y plane, should possess regions where the LDOS is equal for both modes.  This implies that the QD would couple equally to both, such that there is a one-to-one correspondence between the QD Bloch sphere states, and the Poincar\'{e}-like states of the near-field.  It is this criterion that is almost completely unstudied \cite{Larque:H1_cavities_photon_pairs:NJP}.

We will demonstrate methods to determine whether the criteria above are met, and use as examples the well-known L3 and H1 PhCC cavities.  We will show that polarization-degenerate coupling to an L3 cavity mode is not possible, whilst a small region in the exact center of the H1 cavity shows polarization-independent emission.

The layout of this paper is as follows: In Section \ref{sec:Purcell_Enhancement}, we will discuss both classical and semiclassical techniques to find an expression for a position and orientation dependent response of the QD to the cavity field. In Section \ref{sec:Properties_of_Investigated_Cavities}, we will briefly describe the properties, such as dimensions, modes, wavelengths and Q factors, of the H1 and L3 cavities we will be investigating. We will then apply this technique in Section \ref{sec:Enhancement_L3_H1} and show the differences between the two types of cavities.  Note that in this paper we deal with the immediate near-field inside the cavity only.

\section{Modelling the light-matter interaction} \label{sec:Purcell_Enhancement}
Modification of the spontaneous emission lifetime of an excited state of a quantum emitter, often known as the Purcell effect \cite{Purcell:Purcell_effect:PhysRev}, has different interpretations in classical and quantum physics, both of which are worth considering here. From a classical point of view, one may consider the system as a current source dissipating energy in the form of electromagnetic waves. The waves are scattered by the surrounding structure and arrive back at the source, stimulating it to dissipate more energy. From a quantum physical point of view, spontaneous emission is caused by stimulated emission from the vacuum field. By placing an atom in an optical cavity, where the vacuum field is altered, the spontaneous emission of the atom can either be enhanced or inhibited.

In this section we will first show different well-established methods, both classical and semiclassical, to describe the basics of the light-matter interaction.

\subsection{Theory} \label{sec:Theory_Purcell}
The most common approach to this problem is a semi-classical method using Fermi's golden rule \cite{Novotny:Nano_Optics, Benistry:Confined_Photon_Systems, Gerard:Strong_Purcell_Quantum_Boxes}. Fermi's golden rule uses time-dependent perturbation theory to find an expression for the decay rate from an initial state $\ket{i}$ to a final state $\ket{f}$:
\begin{equation}
    \gamma=\frac{2\pi}{\hbar^2}|\bra{f}\bm{H}_i\ket{i}|^2\rho(\omega)
    \label{eq:Fermis_golden_rule}
\end{equation}
where $\bm{H}_i=\bm{\mu}\cdotp\bm{\mathcal{E}}(\bm{r}_0)$ is the interaction Hamiltonian in the dipole approximation, $\bm{\mathcal{E}}(\bm{r}_0)$ the vacuum field at the source location, $\bm{\mu}$ the dipole moment and $\rho(\omega)$ the density of available optical states. The well-known Purcell factor is then given by the ratio of the decay rate in the cavity to that in free space ($\gamma_0$) plus the enhancement due to leaky modes of the photonic crystal and cavity ($F_{pc}$)\cite{Benistry:Confined_Photon_Systems}:
\begin{equation}
    Fp=\frac{\gamma}{\gamma_0}=\frac{3Q(\lambda_c/n)^3}{4\pi^2V_m}\eta^2\frac{1}{1+4Q^2[(\omega/\omega_c)-1]^2}
    \label{eq:purcell_factor_extended}
\end{equation}
where $Q$ is the quality factor of the cavity, $\lambda_c$ the resonant cavity wavelength in free space, $n$ the refractive index of the medium, $V_m$ the modal volume as given by Eq.~(\ref{eq:modal_volume}), $\eta$ the spatial mismatch term given by $\frac{|\bm{\mu}\cdotp\bm{E}(\bm{r}_0)|}{|\bm{\mu}||\bm{E}_{max}|}$ where $\bm{E}(\bm{r}_0)$ is the local electric field at the source location and $\bm{E}_{max}$ is the maximum field strength in the cavity. We point out that $\bm{E}(\bm{r})$ is a classical field that obeys Maxwell's equations. The Lorentzian term $\frac{1}{1+4Q^2[(\omega/\omega_c)-1]^2}$ describes the spectral mismatch of the dipole with the cavity resonance. When the dipole is perfectly aligned with an anti-node of the cavity field and spectrally with the cavity mode, Eq.~(\ref{eq:purcell_factor_extended}) reduces to:
\begin{equation}
    Fp_{max}=\frac{3Q(\lambda_c/n)^3}{4\pi^2V_{m}}
    \label{eq:purcell_factor_alligned}
\end{equation}
The modal volume, $V_m$, is determined using \cite{Kristensen:mode_volume:OL}:
\begin{equation}
    \frac{1}{V_m}=\real \left\{\frac{1}{v_m}\right\},   v_m= \frac{\langle\langle\bm{f}_c|\bm{f}_c\rangle\rangle}{\epsilon(\bm{r}_c)\bm{f}_c^2(\bm{r}_c)}
    \label{eq:modal_volume}
\end{equation}
where $\bm{f}_c$ is the cavity mode calculated from a wave equation with outgoing wave boundary conditions and $\bm{r}_c$ denotes the location of the cavity field antinode. The norm $\langle\langle\bm{f}_x|\bm{f}_y\rangle\rangle$ is defined as:
\begin{equation}
    \langle\langle\bm{f}_x|\bm{f}_y\rangle\rangle=\lim_{V \rightarrow \infty}\int_V \epsilon(\bm{r})\bm{f}_x(\bm{r})\cdot\bm{f}_y(\bm{r}) d\bm{r} + i\frac{\sqrt{\epsilon}c}{\omega_x+\omega_y}\int_{\delta V}\bm{f}_x(\bm{r})\cdot\bm{f}_y(\bm{r}) d\bm{r} = \delta_{x,y}
    \label{eq:modal_volume_norm}
\end{equation}
where $\omega_{x,y}$ are complex frequencies due to the boundary conditions used and $\delta V$ is the border of the volume $V$.
It should be noted that Fermi's golden rule is only valid in the case that the spectral linewidth of the dipole is much smaller than the cavity linewidth \cite{Gerard:Strong_Purcell_Quantum_Boxes} and in the weak coupling regime. Although these two conditions are met in the scope of this work, we will not use the framework of Fermi's golden rule for the analysis in this paper. The complete analysis can however be performed within this framework (see analysis for Fig.~\ref{fig:angles_excited_modes}).

Alternatively, we may calculate the Purcell enhancement from the ratio of energy dissipation by an infinitely small current source in the photonic structure to that in a bulk material \cite{Novotny:Nano_Optics}.
The rate of energy dissipation by a current source, $dW/dT$, assuming a harmonic time dependence for both the current source and electric field, is given by:
\begin{equation}
    \frac{dW}{dt} = -\frac{1}{2}\int_{V_s}\real\left\{ \bm{j}^*\cdotp\bm{E}\right\}d\bm{r}
    \label{eq:energy_dissipation_current}
\end{equation}
where $V_s$ is the source volume, $\bm{j}$ the source current and $\bm{E}$ the electric field vector. Following \cite{Novotny:Nano_Optics}, we can write the source current as $\bm{j}(\bm{r})=i\omega\bm{\mu}\delta[\bm{r}-\bm{r}_0]$, where $\bm{\mu}$ is the dipole moment of the source and $\bm{r}_0$ is the center of the charge distribution of the dipole. Using this we can rewrite Eq.~(\ref{eq:energy_dissipation_current}) as:
\begin{equation}
    \frac{dW}{dt} = -\frac{\omega}{2}\imag\left\{ \bm{\mu}^*\cdotp\bm{E}(\bm{r}_0)\right\}
    \label{eq:energy_dissipation_dipole}
\end{equation}
Alternatively, we may write the electric field created by the source in terms of its dyadic Green's function (DGF) \cite{Novotny:Nano_Optics}:
\begin{equation}
    \bm{E}(\bm{r})=\omega^2\mu\mu_0\bm{G}(\bm{r},\bm{r}_0;\omega)\bm{\mu}
    \label{eq:field_greens_function}
\end{equation}
where $\mu$ is the relative magnetic permeability, $\mu_0$ the vacuum permeability and $\bm{G}(\bm{r},\bm{r_0};\omega)$ the DGF. It is defined as:
\begin{equation}
    \nabla\times\nabla\times\bm{G}(\bm{r},\bm{r_0};\omega)- \frac{\omega^2}{c^2}\epsilon(\bm{r})\bm{G}(\bm{r},\bm{r_0};\omega) = \bm{I}\delta(\bm{r}-\bm{r_0})
    \label{eq:greens_def}
\end{equation}
where $c$ is the vacuum speed of light, $\epsilon$ the relative electric permittivity and $\bm{I}$ is the unit dyad. Here, $\bm{G}$ both includes transverse and longitudinal contributions e.g. $\bm{G}=\bm{G}^T+\bm{G}^L$. Using Eqs.~(\ref{eq:energy_dissipation_dipole}) and (\ref{eq:field_greens_function}) we can rewrite the energy dissipation of the current source as:
\begin{equation}
    \frac{dW}{dt} = -\frac{\omega^3|\bm{\mu}|^2}{2c^2\epsilon\epsilon_0} \imag\left\{ \hat{\bm{\mu}}^{*}\cdotp\bm{G}(\bm{r}_0,\bm{r}_0;\omega)\cdotp\hat{\bm{\mu}} \right\}
    \label{eq:energy_dissipation_greens}
\end{equation}
where $\hat{\bm{\mu}}$ is the normalized dipole moment and $\epsilon_0$ the vacuum permittivity. Note that here we slightly differ from \cite{Novotny:Nano_Optics} in order to take complex dipoles into account. We can then use Eq.~(\ref{eq:energy_dissipation_greens}) to calculate the energy dissipation of the source. The ratio between the power dissipation of the source in the cavity and in a bulk material is then given by:
\begin{equation}
    Fp(\bm{r}_0)=\frac{\imag\left\{ \hat{\bm{\mu}}^{*}\cdotp\bm{G}(\bm{r}_0,\bm{r}_0;\omega)\cdotp\hat{\bm{\mu}}\right\}}{\imag\left\{ \hat{\bm{\mu}}^{*}\cdotp\bm{G_0}(\bm{r}_0,\bm{r}_0;\omega)\cdotp\hat{\bm{\mu}}\right\}}
    \label{eq:purcell_ratio_greens}
\end{equation}
where $\imag\left\{\bm{G_0}(\bm{r}_0,\bm{r}_0;\omega)\right\}=\omega\sqrt{\epsilon}/6\pi c$ is the DGF in a bulk material. The Purcell factor is the normalized equivalent of the semi-classical LDOS, where the LDOS is also proportional to the imaginary part of the DGF \cite{Vos:Orientation_dependent_spontaneous_emission:PRA}. We can now introduce the transverse DGF for an arbitrary cavity structure at the dipole location following \cite{Cowan:Mode_expansion:PRE, Hughes:QD_cav_wav_strong:PRB}:
\begin{equation}
    \bm{G}^T(\bm{r}_0,\bm{r}_0;\omega)=\frac{c^{2}\bm{e}_{c}(\bm{r}_0)\otimes\bm{e}_{c}^{*}(\bm{r}_0)}{\omega_{c}^{2}-\omega^{2}-i\omega\omega_{c}/Q}
    \label{eq:greens_cavity_expantion}
\end{equation}
where $\omega_{c}$ is the cavity resonant frequency and $\bm{e}_{c}$ is the normalized cavity mode function given by:
\begin{equation}
    \bm{e}_{c}(\bm{r})=\bm{E}_{c}(\bm{r})/\sqrt{V_{m}}\maxm({\sqrt{\epsilon}\bm{E}_{c}(\bm{r})})
    \label{eq:greens_mode_func_def}
\end{equation}
Because we only consider the transverse part of the DGF here, we do not take coupling of the QD with leaky modes into account. Combining Eqs.~(\ref{eq:purcell_ratio_greens}), (\ref{eq:greens_cavity_expantion}) and the GDF in bulk material finally results in:
\begin{equation}
    Fp(\bm{r}_0)=\frac{6\pi c^{3}}{\omega\sqrt{\epsilon}}
    \imag\left\{ \hat{\bm{\mu}}^{*}\cdotp \frac{\bm{e}_{c}(\bm{r}_0)\otimes\bm{e}_{c}^{*}(\bm{r}_0)}{\omega_{c}^{2}-\omega^{2}-i\omega\omega_{c}/Q}\cdotp\hat{\bm{\mu}}\right\}
    \label{eq:greens_final}
\end{equation}

\subsection{Modelling of the polarization-dependent QD-cavity interaction} \label{sec:QD_model}
Equations~(\ref{eq:energy_dissipation_current}-\ref{eq:greens_final}) give a general description of how dipole emission is modified in a cavity system due to the change in light-matter interaction. These equations do not directly account for the time-dependent decay dynamics of QDs but merely take a dipole moment into account. We are thus making the assumption that the QD in the cavity may be modeled as a superposition of dipole emitters with particular orientations in the x-y plane. Taking the trion transition in a QD as an example, a trion in the superposition state $\sqrt{1/2}(|\uparrow,\downarrow;\Uparrow\rangle \pm |\uparrow,\downarrow ; \Downarrow \rangle)$ emits an H(V) photon, leaving an electron in superposition state $\sqrt{1/2}(|\uparrow\rangle \pm |\downarrow \rangle)$, corresponds to a dipole emitter oriented along the $x(y)$-axis, respectively. Indeed, one may describe any arbitrary transition on the spin Bloch sphere as two dipole emitters oscillating with a given relative phase.  For example, a spin $|\uparrow\rangle$ (coupling to $|L\rangle$ circular polarization) may be modelled by using two dipole emitters that have a relative phase of $+\pi/2$. However, when the strong coupling regime or QD-QD interactions are of interest, the time dependent decay dynamics of the QD have to be taken into account \cite{Hughes:Single_QD_PhC:OptLet,Hughes:Double_QD_PhC:PRL}.

To calculate how a dipole emitter with a particular orientation and position in the cavity will couple to the cavity modes, one may use the Green's function method described in \cite{Kristensen:2QDs_in_PhC:PRB}. As we will demonstrate, using the Green's function method is more straightforward than using Eq. (\ref{eq:purcell_factor_extended}), as the situation is not trivial for situations such as degenerate cavities. The Green's function method describes, for a particular QD orientation and direction, the complete coupling to the electromagnetic environment. The approach in \cite{Kristensen:2QDs_in_PhC:PRB} also takes strong coupling, coupling of several emitters and several mode frequencies into account. In the following section we show, using a similar Green's function method, that for a spin-photon interface, one must split the Green's function into two degenerate, orthogonal states that have a one-to-one correspondence with the QD spin-transition. We show that an (x,y) ``map'' of the orthogonal states is beneficial in identifying optimal positions in the cavity, and whether cavities are at all suitable as a spin-photon interface.

\section{Properties of the investigated cavities} \label{sec:Properties_of_Investigated_Cavities}
There are several types of optical microcavities used for coupling QDs to a controlled photonic environment. Several designs have demonstrated strong coupling, such as circular\cite{Reithmaier:Strong_Coupling:Nature} and elliptical\cite{Reizenstein:SCpolarization:PRB} micropillars, photonic crystals\cite{Yoshie:Strong_Coupling:Nature,Ota:H1SC:2009}, and microdisks\cite{Peter:microdisc:2005}. The methods we apply below (identifying the photonic superposition states, and the correspondence between the spin and superposition states) will apply to any lossless dielectric cavity system. We choose to study photonic crystals in detail here.  This is because these represent the most complex case: a circular micropillar or microdisk has a radially symmetric LDOS which is relatively easy to analyze in terms of its polarization.  Even upon lowering the symmetry (for example in an elliptical pillar) the LDOS varies more slowly spatially (nevertheless, the polarization-dependent coupling of a QD to these now-polarized modes is highly non-trivial\cite{Reizenstein:SCpolarization:PRB}).  Photonic crystal cavities, on the other hand, do not show rotational symmetry of their degenerate modes, and show a strong spatial dependence in the polarization of the LDOS, as we demonstrate. The lack of radial symmetry makes identification of suitable superposition states non-trivial, as these states also do not show rotational symmetry.  More recent focus on plasmonic-QD structures (e.g. \cite{Gan:plasmonsource:2012,Biagioni:Cross_antenna:2009,Chen:QDWG:2010}) suggests that ultrasmall ($<< (n\lambda)^3$) metal plasmonic cavities and antenna structures may also be used, again with strong coupling theoretically possible.  However, these metal structures are dispersive. This means that the techniques used here do not directly translate to plasmonics, or lossy systems, as the mode expansion is not as straightforward as described here \cite{Sondergard:Greens_plasmonics:PRB}. However, we hope to address this in future work.

\begin{figure}[htbp]
    \centering\includegraphics[width=0.6 \columnwidth ]{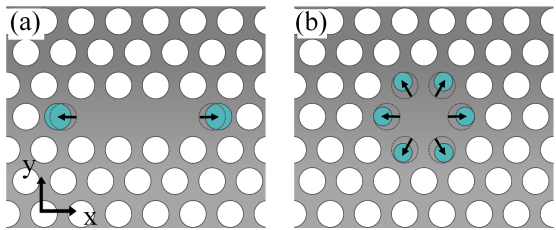}
    \caption {Schematic representation of the (a) L3 and (b) H1 photonic crystal cavities. The teal circle indicates the altered
    holes surrounding the cavity. The dotted circles indicate their unaltered size and position.}
    \label{fig:cavity_layout}
\end{figure}

By way of demonstration, we will investigate two PhCC designs, both shown in Fig.~\ref{fig:cavity_layout}.  The L3 cavity in Fig.~\ref{fig:cavity_layout}(a) consists of a line of three missing holes.  The dimensions of the L3 cavity we used are based on those presented in \cite{Chalcraft:L3_mode_structure:APL}, where the end holes of the cavity are displaced to obtain a smoother transition between the defect and photonic crystal, resulting in a higher Q-factor. The L3 cavity has the following parameters: a hole radius of $r=0.28a$, where a is the lattice constant, a slab thickness of $d=0.58a$ and an end hole displacement of $s=0.15a$. This cavity design is well-studied, and was the first PhCC design to produce strong QD-cavity coupling \cite{Yoshie:Strong_Coupling:Nature}. This cavity design, however, is not known to show any degenerate mode and, experimentally, all modes measured from this cavity are strongly polarizing \cite{Oulton:Polarized_L3_emission:OptExpr, Chalcraft:L3_mode_structure:APL}. We therefore already do not expect that this design is suitable for polarization-degenerate coupling.

The H1 cavity, based on the design in \cite{Kim:Vertical_Beaming:PRB}, on the other hand, shows six fold rotational symmetry, $C_{6v}$, giving rise to an energy degenerate fundamental mode \cite{Painter:Local_Defects:PRB}.  The exact design is shown in Fig.~\ref{fig:cavity_layout}(b).  The structure has a hole radius of $r=0.35a$, a slab thickness of $d=0.5a$, an end hole displacement of $s=0.10a$, the radius of the end holes is $r_d=0.25a$. All cavities are created in a GaAs slab membrane with a refractive index of $n=3.4$. Because the photonic crystal is realized in a slab, only a bandgap for TE-modes is present and thus only TE-modes are confined by the PhCCs, meaning the modes only have $H_z$, $E_x$ and $E_y$ field components \cite{Joannopoulos:Molding_the_Flow_of_Light}.

We simulate the structures with the finite difference time domain (FDTD) method \cite{Taflove:FDTDbook}, using a modified version of an opensource software package \cite{Oskooi:MEEP}.  A lattice of $13a\times9a$ and $11a\times11a$ for the L3 and H1 cavity respectively is used, we chose a resolution of 90 grid points per lattice period. %, and the QD simulated using a current source acting as the dipole emitter, occupying a single unit cell. %The position of the current source allows one to simulate the effect of QD position.
The first step is to find the frequency of any high-Q modes with which the QD interacts.  This is performed by producing a spectrally wide current pulse, a normalized width of $\Delta\omega_c/\omega_c=0.1$ is used, allowing a wide range of modes to be excited simultaneously.  The modes and their Q-factors are identified using Fourier transform and a harmonic inversion technique \cite{Mahdelshtam:Harminv_ref}. We then proceed to extract the amplitude and phase information of the mode by outputting time harmonic fields at the center of the mode resonance, using a run time Fourier transform.

\begin{figure}[htbp]
    \centering\includegraphics[width=1 \columnwidth ]{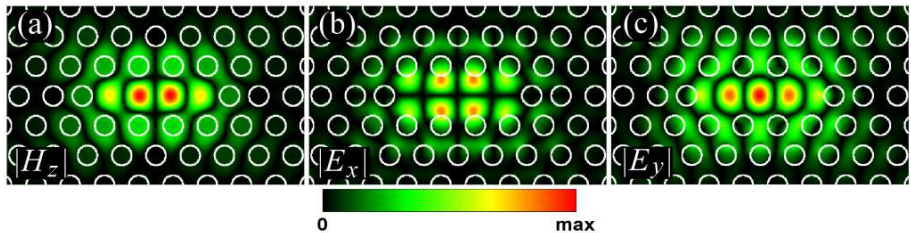}
    \caption {Cavity field distributions of the L3 $[^{-}_{-}1]$ mode (a) $|H_z|$, (b) $|E_x|$ and (c) $|E_y|$.}
    \label{fig:L3_Cavity_modes}
\end{figure}

\begin{figure}[htbp]
    \centering\includegraphics[width=1 \columnwidth ]{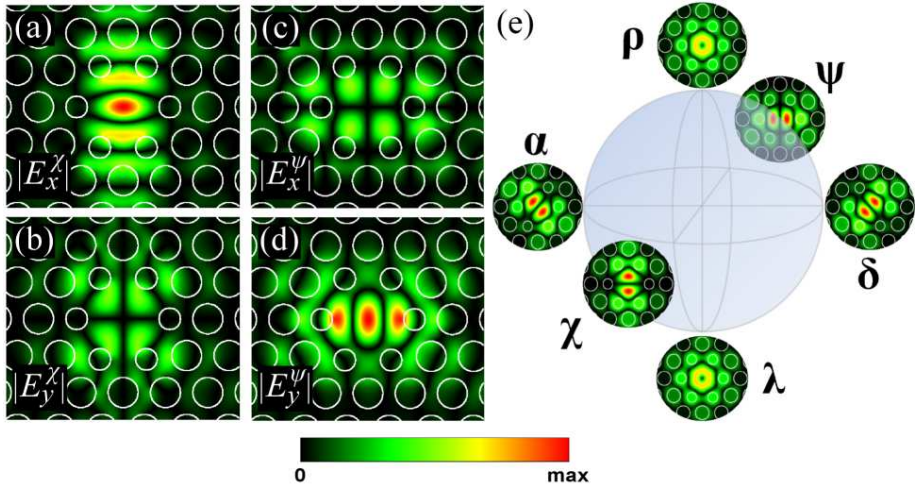}
    \caption { Cavity field distributions for the H1 cavity (a) $|E_x|$ and (b) $|E_y|$ field distributions of the $\chi$ dipole mode and (c) $|E_x|$ and (d) $|E_y|$ field distributions of the $\psi$ dipole mode. (e) Poincar\'{e}--like sphere with different orientations of the $|H_z|$ component of the H1 dipole mode.}
    \label{fig:H1_Cavity_modes}
\end{figure}

An overview of the L3 cavity modes and their properties can be found in \cite{Chalcraft:L3_mode_structure:APL} and in \cite{Kim:Vertical_Beaming:PRB} for the H1 cavity. In this paper we focus on the fundamental $[^{-}_{-}1]$ mode in the L3 cavity (as shown in Fig.~\ref{fig:L3_Cavity_modes}) which is of most interest due to its high Q-factor and the dipole mode in the H1 cavity (as shown in Fig.~\ref{fig:H1_Cavity_modes}) which is of interest due to its double degeneracy.
The Q-factors found in our simulations, 7600 for the fundamental L3 mode and 1300 for the H1 dipole mode, are relatively low compared to the 47,000 and 15,000 mentioned in \cite{Chalcraft:L3_mode_structure:APL, Kim:Vertical_Beaming:PRB} for the L3 and H1 fundamental modes. This is due to the limited size of the PhC. By adding more periods to the photonic crystal an increase in the Q-factor can be obtained. However, considering the large resolution of 90 grid points per lattice period used to simulate these structures, the computational cell needed to be limited to keep computation feasible. After testing with larger structures in our simulations, we are confident that the reduced size of the PhC does not noticeably affect the polarization properties we study in this paper.

Figure~\ref{fig:L3_Cavity_modes} shows the cavity electric and magnetic field distributions of the L3 cavity, generated by placing a dipole source in the center of the cavity.  We extract the field distributions at the peak wavelength of the fundamental mode and show that it agrees well with calculations obtained using guided mode expansion techniques \cite{Chalcraft:L3_mode_structure:APL}.

We notice from Fig.~\ref{fig:L3_Cavity_modes}(b,c) that the fundamental mode contains both $x$ and $y$ components.  However, their spatial profile differs greatly, which strongly affects the coupling of the QD to the mode.  For example a QD placed directly in the center will only couple if it has an $y$ component to its dipole, as the $E_x$ component at the centre is zero.  Likewise, there are regions of the cavity where the QD dipole should have an $x$ component to couple to the cavity, as the $E_y$ component is zero.  It already becomes clear here that polarization information is not preserved when a QD couples to the cavity.  The orientation of the QD's effective dipole will determine the rate of coupling to the cavity, as given by Eq.~(\ref{eq:greens_final}).  However, all photons emitted by the QD will emit into the mode, which has its own, independent polarization profile, and none of the original spin/polarization information remains.

Thus it would already appear that the L3 cavity is unpromising for spin-based QD applications. More promising is the H1 cavity.  Unlike the L3 cavity, the H1 cavity has two sets of doubly degenerate modes, the ``dipole'' and ``quadrupole'' modes, as well as two non degenerate modes, the ``hexapole'' and ``monopole'' modes \cite{Kim:Vertical_Beaming:PRB}. Figure~\ref{fig:H1_Cavity_modes}(a-d) shows the $|E_x|$ and $|E_y|$ components of the two dipole modes, as calculated by placing an $x$ and $y$ oriented dipole in the center, for the mode in Fig.~\ref{fig:H1_Cavity_modes}(a,b) and Fig.~\ref{fig:H1_Cavity_modes}(c,d) respectively.

All these modes, just as those in the L3 cavity, are eigenmodes of the PhCC. Eigenmodes have the property that a linear combination of the two modes is also a valid mode. As we will see, this makes these modes much more viable for use in spin-based QD applications.
Geometry effects in environment-dependent coupling is discussed in detail in \cite{Vos:Orientation_dependent_spontaneous_emission:PRA}, where the spontaneous emission rate of a dipole in the presence of an arbitrarily-shaped nano-photonic environment is analyzed. Counterintuitively, there is no simple relationship between the symmetry of the photonic environment and the directional dependence of the Purcell enhancement.  For example, as the H1 cavity has a six-fold degeneracy, one might expect a six-fold degeneracy in the DGF, and therefore the Purcell enhancement.  However, because the DGF may always be diagonalized, it will have three eigenvalues $(g_a,g_b,g_c)$, that correspond to three eigenmodes of the system. In our system however, because we are dealing with TE-modes only, the third eigenvalue will always be zero and can be neglected. When considering Purcell enhancement in the cavity, these eigenmodes correspond to two principle dipole orientations with maximum and minimum enhancement, ${\Gamma_{max}, \Gamma_{min} }$.  The emission rate does not inherit the symmetry of the cavity system.  Note also that the DGF is dependent both on the nanophotonic environment \emph{and} the position of the dipole emitter.

In our system, we wish to find cavity designs which have QD positions where $\Gamma_{max} = \Gamma_{min}$.  From the analysis above, this would mean that the two eigenvalues $(g_a,g_b)$ are equal. From Eq.~(\ref{eq:greens_cavity_expantion}) it can be easily shown that the eigenvalues of the GDF can only be equal in the case that two orthogonal modes are present and overlap spectrally, the total GDF is then simply the sum of the GDF of the orthogonal modes, e.g:
\begin{equation}
    \bm{G}^T(\bm{r}_0,\bm{r}_0;\omega)=c^{2}\sum_{n\in1,2}\frac{\bm{e}_{cn}(\bm{r}_0)\otimes\bm{e}_{cn}^{*}(\bm{r}_0)}{\omega_{cn}^{2}-\omega^{2}-i\omega\omega_{cn}/Q_n}
    \label{eq:greens_summation}
\end{equation}
where $e_{cn}$ is $e_c$ of the $n^{\text{th}}$ cavity mode. The second requirement, finding optimal QD positions within the $x,y$ plane is more difficult, and we will discuss this in Section \ref{sec:Enhancement_L3_H1}.  However, the first requirement that the eigenvalues are the same naturally implies that two orthogonal and degenerate modes must be present.  We therefore need to first determine whether this is the case for the particular cavity in question.

The dipole mode, which is the only one we consider in this paper, may be described in many different orthogonal basis vector fields. Two vector fields are said to be orthogonal if their inner product is zero. The inner product between two vector fields is given by:

\begin{equation}
    (\bm{H}^{a},\bm{H}^{b})=\int_V\bm{H}^{a^*}(\bm{r})\cdotp \bm{H}^{b}(\bm{r})d^3\bm{r}
    \label{eq:vector_field_innerproduct}
\end{equation}

With this knowledge we can construct different sets of basis vector fields that describe the dipole mode of the H1 cavity.  We choose the H1 dipole mode as it has been studied fairly widely previously \cite{Larque:H1_cavities_photon_pairs:NJP, Painter:Local_Defects:PRB}. These studies have shown that two degenerate dipole modes exist in the H1 cavities.  These may be identified by FDTD simulations using an $x$ and $y$ polarized dipole emitter in the center of the cavity. Figure~\ref{fig:H1_Cavity_modes}(a-d) show the $|E^{\chi}_x(x,y)|$, $|E^{\chi}_y(x,y)|$ field distributions resulting from an $x$-dipole, and the $|E^{\psi}_x(x,y)|$ and $|E^{\psi}_y(x,y)|$ field distributions resulting from a $y$-dipole.  To avoid confusion with the field components, we have named the different orientations of the mode with a Greek letter. For instance the $\chi$-mode, noted in terms of its magnetic vector field as $\bm{H}^{\chi}$, is the mode that can be excited with a x-dipole in the center of the cavity, while the $\psi$-mode is excited with a y-dipole in the center of the cavity.

Just as diagonally and anti-diagonally-polarized photon states may be constructed as the superposition of an $x$ and $y$-polarized photon, so the ``$\delta$'' (diagonal) and ``$\alpha$'' (antidiagonal) modes may be constructed with the superposition of the $\chi$-mode and $\psi$-mode eg, $\bm{H}^{\delta / \alpha}(x,y)= \left(\bm{H}^{\chi}(x,y)\pm\bm{H}^{\psi}(x,y)\right)/\sqrt{2}$. These modes are excited by diagonal and anti-diagonal dipoles in the center of the cavity. Note that one does not need to re-run the FDTD simultion, as it is identical to adding the vector fields.  Similarly, the $\rho$ and $\lambda$ modes are formed from a phase delayed superposition of the $\chi$- and $\psi$-modes: $\bm{H}^{\rho / \lambda}(x,y)= \left(\bm{H}^{\chi}(x,y)\pm  i\bm{H}^{\psi}(x,y)\right)/\sqrt{2}$ and can be excited using a right or left circular polarizing dipole. One may form a Poincar\'{e}-like sphere on which all possible states of the cavity mode can be mapped. The modes denoted by the Greek letters $\chi, \psi, \delta, \alpha, \rho, \lambda$, are near-field equivalents of the far-field paraxial $x,y,d,a,R,L$ polarization states on the conventional Poincar\'{e} sphere, respectively.  Figure~\ref{fig:H1_Cavity_modes}(a) shows a Poincar\'{e}-like sphere, which shows the $|H_z(x,y)|$ field distribution for the bases discussed above. Just as the Poincar\'{e} sphere's surface represents the complete set of possible photon polarizations in the far-field, this sphere represents the equivalent for the near-field in the cavity and has a one-to-one correspondence to the spin Bloch sphere. The ability to form a complete set of states on the Poincar\'{e}-like sphere is crucial for quantum information applications. It must be possible to create a set of photonic cavity states that allow a full set of superposition states to be contained with full fidelity in the cavity.

\section{Enhancement in L3 and H1 cavities} \label{sec:Enhancement_L3_H1}
So far we have examined L3 and H1 cavity modes and shown that for the energy-degenerate H1 cavity, it is possible to generate a full set of photonic modes that would represent superposition states.  However, the position of the QD determines how much of each component is excited.  We examine this in this section.  Using the results from Section \ref{sec:Purcell_Enhancement}, we are now able to look at the effects on the Purcell enhancements of an emitter in a PhCC. We will start with the $[^-_-1]$ mode of the L3 cavity followed by the dipole mode of the H1 cavity.

\subsection{Enhancement in L3 cavities} \label{sec:Enhancement_L3}
\begin{figure}[htbp]
    \centering\includegraphics[width=1 \columnwidth ]{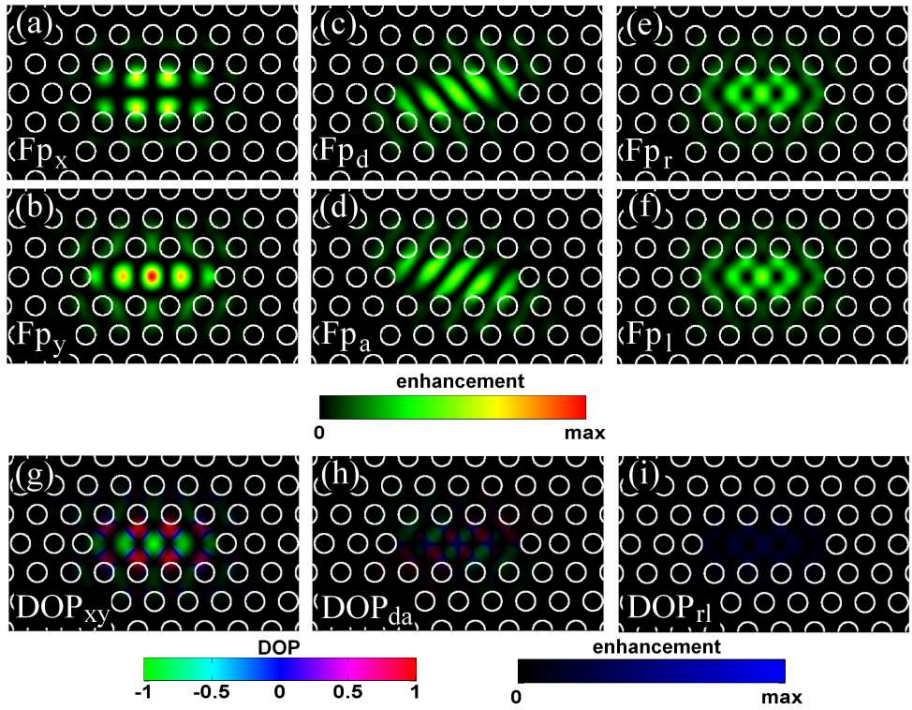}
    \caption {Spatial map of the Purcell factor for L3 cavities for (a) a horizontal, x, dipole (b) a vertical, y, dipole (c) a diagonal, d, dipole (d) an anti-diagonal, a, dipole (e) a right, r, dipole and (f) a left, l, dipole. (g) Degree of polarization for an x and y dipole (h) degree of polarization for the d and a dipole and (i) degree of polarization for a r and l dipole (note that the DOP in (i) is almost zero).}
    \label{fig:L3_enhancement}
\end{figure}

From Eq.~(\ref{eq:greens_final}) we know that the enhancement is proportional to the square of the electric field. The cavity field of the $[^-_-1]$ mode can be seen in Fig.~\ref{fig:L3_Cavity_modes}(b-c). We now plot the enhancement for both x and y oriented dipoles ($Fp_x(x,y)$) and ($Fp_y(x,y)$) as a function of position, as shown in Fig.~\ref{fig:L3_enhancement}(a-b) assuming that the QD transition is on resonance with the cavity mode. We can also plot the enhancement for diagonal (c) and anti-diagonal (d) right handed (e) or left handed (f) oriented dipole sources by rotating the $E_x(x,y)$ and $E_y(x,y)$ fields to their $E_d(x,y)$, $E_a(x,y)$, $E_r(x,y)$ and $E_l(x,y)$ counterparts, i.e. $E_{d/a}(x,y) = (E_x(x,y) \pm E_y(x,y))/\sqrt{2}$ and $E_{r/l}(x,y) = (E_x(x,y) \pm iE_y(x,y))/\sqrt{2}$ and calculating the GDF using Eq.~(\ref{eq:greens_final}) in these bases or by simply rotating the dipole moment. From these six graphs we can already clearly tell that the cavity-dipole coupling is highly direction dependent for linear sources. For instance at the peak of the enhancement Eq.~(\ref{eq:purcell_factor_alligned}) tells us the Purcell factor is about 750. This point is exactly in the center of the cavity where there is a node in the $E_x$ field component and an antinode at the $E_y$ component. This means that the $x$ component of a dipole will not couple to the cavity mode and, depending on the $F_{pc}$ factor in Eq.~(\ref{eq:purcell_factor_extended}), emission might be completely suppressed. On the other hand, for a circular polarized source the Purcell enhancements are identical for both left and right circular polarization.

To gain a better understanding of the polarizing properties of the cavity, we plot the degree of polarization (DOP):
\begin{equation}
    DOP_{i,j}(x,y) = \frac{Fp_i(x,y) - Fp_j(x,y)}{Fp_i(x,y) + Fp_j(x,y)}
    \label{eq:dop}
\end{equation}
where $i,j$ are orthogonal bases ($x,y;d,a;r,l$), in Fig.~\ref{fig:L3_enhancement}(g-i). In these figures the hue shows the polarization eg, red means fully x-polarized, green means fully y-polarized and blue is non-polarizing (i.e. equal coupling to both polarizations). The intensity of the color shows the Purcell enhancement for that region scaled with an enhancement assuming perfect alignment with the field e.g. scaled by $Fp(x,y) = Fp_x(x,y) + Fp_y(x,y)$.
From these plots it is clear that for each position within the cavity, a preferred dipole polarization exists.  Blue areas, indicating an equal coupling, only occur where there are nodes in intensity. For no point in the cavity is there a region of non-zero intensity which couples equally to all polarizations.

\begin{figure}[htbp]
    \centering\includegraphics[width=1 \columnwidth ]{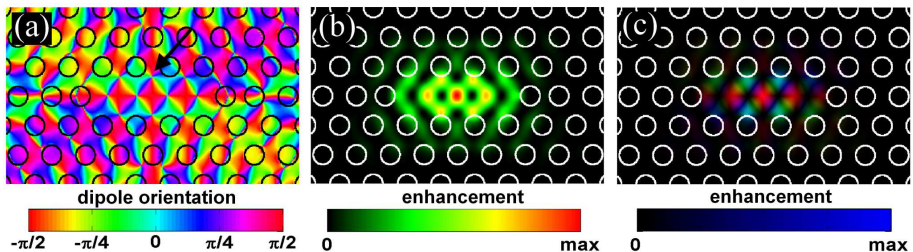}
    \caption {(a) Angle of the dipole for which maximum enhancement is achieved. (b) Enhancement of the dipole when it is oriented at the optimal angle shown in (a).(c) Combination of plots (a) and (b), where the angles in (a) have been normalised to intensity using (b). This indicates that all areas of degenerate coupling do not show any enhanced emission.}
    \label{fig:L3_max_enhancement}
\end{figure}

Alternatively, we can calculate the angle of the dipole that gives the maximum enhancement. To do this we define the dipole moment as $\bm{\mu} = [\cos(\theta) ; \sin(\theta) ]$, where $\theta = 0$ corresponds to ``H'' polarization, and $\theta = 90^0$ corresponds to ``V'' polarization, and we apply this to Eq.~(\ref{eq:greens_final}). By differentiating this equation with respect to $\theta$ and setting it to zero we can find the maximum enhancement angle $\theta_{max}$:
\begin{equation}
    \theta_{max}(\bm{r}_0) = \frac{1}{2}\arctan\left(\frac{\imag\{G_{1,2}(\bm{r}_0,\bm{r}_0) + G_{2,1}(\bm{r}_0,\bm{r}_0)\}}{\imag\{G_{1,1}(\bm{r}_0,\bm{r}_0) + G_{2,2}(\bm{r}_0,\bm{r}_0)\}}\right)
    \label{eq:maximum_enhancement_angle}
\end{equation}
where $G_{i,j}$ is the matrix element of the GDF on the $i$-th row and $j$-th column. Figure~\ref{fig:L3_max_enhancement}(a) shows the maximum enhancement angle for the L3-cavity, while Fig.~\ref{fig:L3_max_enhancement}(b) shows the maximum possible enhancement $F_{max}(x,y)$. The center of the cavity shows a maximum enhancement for a dipole at an angle of $\pi/2$ which confirms the plots shown in Fig.~\ref{fig:L3_enhancement}(a-f). A few singularities can also be seen: these are points where the angle is undefined. One such point is indicated by a black arrow. These singularities show points in the PhCC where all dipole orientations are enhanced equally, e.g. the eigenvalues of the DGF at this point are equal. This would suggest that a spin-photon interface could be implemented in L3-cavities. However, if we look at the Purcell enhancement for dipoles rotated to the maximum enhancement angles as shown in Fig.~\ref{fig:L3_max_enhancement}(b), we observe there is no enhancement at these points, e.g. the eigenvalues are zero. QDs positioned at these points are not able to emit into the cavity mode and can only emit into leaky modes, this greatly inhibits the spontaneous emission. To visualize this we combine Fig.~\ref{fig:L3_max_enhancement}(a-b) to Fig.~\ref{fig:L3_max_enhancement}(c) where the intensity of the angles in Fig.~\ref{fig:L3_max_enhancement}(a) have been scaled by the enhancement. In this plot the singularities have disappeared and no points of orientation independent coupling can be seen. This behaviour makes the L3 cavity unsuitable for spin-photon interfaces or the entangling photon sources using the biexciton cascade of a QD.

\subsection{Enhancement in H1 cavities}  \label{sec:Enhancement_H1}
Analysis of the polarization-dependent Purcell enhancement of the dipole mode in the H1 cavity is somewhat more complex than in the L3 cavity.  This is due to the presence of two degenerate modes.  As we have discussed, the Purcell enhancement arises from the QD dipole exciting the mode, which, classically, backscatters the light onto the dipole, increasing stimulated emission.  Calculation of this dipole-mode coupling is relatively easy for an L3 cavity, as only one mode is excited. However, for an H1 cavity, the dipole couples in general to two modes.  Thus, whilst the two degenerate modes are orthogonal (as shown in Fig.~\ref{fig:H1_Cavity_modes}), the dipole may couple the two modes, giving rise to more complex backscattering of both modes onto the dipole.

For instance, consider the $\chi$ and $\psi$ modes of the cavity, which were generated in Section \ref{sec:Properties_of_Investigated_Cavities} with a dipole placed in the center, and oriented along $x$ and $y$, respectively (Fig.~\ref{fig:H1_Cavity_modes}(a-d)).  We are able to make this statement because an $x$ dipole will only couple to the $x$ component of any mode in the cavity.  Examining Fig.~\ref{fig:H1_Cavity_modes}(c) more closely, we see that the $\psi$ mode has a node for the $E^{\psi}_x(0,0) = 0$ component in the center. It will therefore not be excited by an $x$-oriented dipole. The same goes for a $y$-oriented dipole: no $\chi$ mode is excited because, as shown in Fig.~\ref{fig:H1_Cavity_modes}(b), $E^{\chi}_y(0,0) = 0$.  We can therefore be sure that the two modes generated are orthogonal to each other.  A dipole oriented along the diagonal, however, would excite both the $\chi$ and $\psi$ modes equally, coupling them together.  Thus the description of the $\chi$ and $\psi$ modes as orthogonal no longer holds when a quantum emitter dipole is added: the dipole-mode system should be considered as a whole. In fact, as we have discussed, the photonic system is better described by a superposition of two orthogonal modes rotated around the ``Poincar\'e -like'' sphere in Fig.~\ref{fig:H1_Cavity_modes}(e).  This is what is described by the DGF.

\begin{figure}[htbp]
    \centering\includegraphics[width=0.85 \columnwidth ]{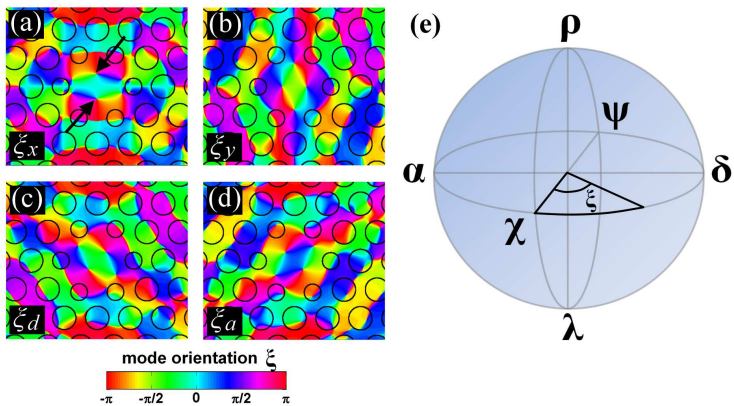}
    \caption {Angles of the excited modes $\bm{E}^{\xi}$ for a source placed at a specific location in the H1 cavity. (a) $\xi_x(\bm{r})$
    (b) $\xi_y(\bm{r})$  (c) $\xi_d(\bm{r})$  (d) $\xi_a(\bm{r})$ and (e) Poincare-like sphere indicating the angle $\xi$ }
    \label{fig:angles_excited_modes}
\end{figure}

In general, one should be able to determine which mode will be excited by a dipole with an arbitrary orientation at any arbitrary position. As we have discussed in Section \ref{sec:Properties_of_Investigated_Cavities}, the H1 cavity modes may be expressed as any pair of orthogonal modes, arising from a linear superposition of the $\chi$ and $\psi$ modes. These modes can be written in terms of the two basis modes and a rotation operator which maps the modes onto two new orthogonal modes on the Poincar\'e-like sphere in Fig.~\ref{fig:H1_Cavity_modes}(a):
\begin{equation}
    \left[ \begin{array}{cc} \bm{E}^{\xi} & \bm{E}^{\xi'} \end{array} \right]=
    \left[ \begin{array}{cc} \cos{\xi} & -\sin{\xi} \\ \sin{\xi} & \cos{\xi} \end{array} \right]
    \left[ \begin{array}{cc} \bm{E}^{\chi} & \bm{E}^{\psi} \end{array} \right]
    \label{eq:mode_rotation}
\end{equation}
where $\bm{E}^{\xi}(x,y)$ and $\bm{E}^{\xi'}(x,y)$ are the two new orthogonal basis modes and $\xi$ is an angle which represents the angle of rotation around the equator of the Poincar\'e-like sphere, as shown in Fig. \ref{fig:angles_excited_modes}(e).
Which basis modes are excited depends on both the source position and dipole orientation.  As we have seen, when the dipole is positioned in the center, rotation of the dipole in the plane by angle $\theta$ excites a superposition of the $\chi$ and $\psi$ modes that is represented by rotation on the Poincar\'e-like sphere of $\xi = \theta$.  This is not the case generally.  An arbitrarily oriented dipole emitter at an arbitrary position will result in a breaking of the symmetry of the system.  Thus, rotation of the dipole on the Bloch sphere equator does not represent a equal rotation on the Poincar\'e-like sphere, and one must find a technique to identify the excited mode. We use Eq.~(\ref{eq:mode_rotation}) to find a rotation angle $\xi$ where one of the field components of the modes $\bm{E}^{\xi'}_p(r) = 0$, where $p$ indicates the field component parallel to the orientation of the dipole. If $\bm{E}^{\xi'}_p(r) = 0$, the dipole emitter at $r$ must be at a node of $\bm{E}^{\xi'}$.  The orthogonal mode of the pair, $\bm{E}^{\xi}$, will give the correct angle of rotation for the system.

If we define the two eigenmodes in this way, we find that the dipole will show no coupling to the $\bm{E}^{\xi'}$ mode, coupling only to the $\bm{E}^{\xi}$ mode.  The angle $\xi$ for an arbitrarily polarized source at position $\bm{r}$ may be expressed as:
\begin{equation}
    \xi_{p}(\bm{r})=\arctan\left( \frac{-E_{p}^{\psi}(\bm{r})}{E_{p}^{\chi}(\bm{r})}\right)
    \label{eq:angle_of_excited_mode}
\end{equation}
The spatial maps $\xi_p(\bm{r})$ for angles of the mode excited by x, y, diagonally and anti diagonally oriented sources are shown in Fig.~\ref{fig:angles_excited_modes}.  These essentially show which type of mode is excited when a dipole of the given orientation (e.g. x) is at position $\bm{r}_0$.  One can see here that the mode excited is dependent on the QD dipole position.  For instance, an $x$ dipole in the center, as shown in Fig.~\ref{fig:angles_excited_modes}(a), excites purely the $\chi$ mode ($\xi_x =0$), whilst in the ``red'' region above the center the $x$-dipole couples purely to the $\psi$ mode ($\xi_x = \pi/2$).  Note that this can give some counterintuitive results: in general, one can make no conclusions about the orientation of the dipole in the cavity simply by examining the mode excited: a vertically polarized dipole emits into the $\psi$ mode if it is in the center, but a horizontally polarized dipole also emits into the $\psi$ mode if it is off-center.
Figure~\ref{fig:angles_excited_modes}(a-d) also show singularities, i.e.  points where the angle is undefined. Two such points have been indicated by arrows in Fig.~\ref{fig:angles_excited_modes}(a). If a dipole is placed at these points, no mode can be excited, which would result in inhibition of the spontaneous emission.  A dipole rotated by $\xi+\pi/2$ placed at this singularity will, however, couple.

Although it is convenient to be able to predict the orientation of the mode depending on the dipole position it is not a necessary part of the analysis when using DGFs. The DGF for the H1 cavity can be calculated using Eq.~(\ref{eq:greens_summation}) and is independent on the basis chosen for the cavity modes. It is, however, a necessity when performing the analysis using Fermi's golden rule.

\begin{figure}[htbp]
    \centering\includegraphics[width=0.75 \columnwidth ]{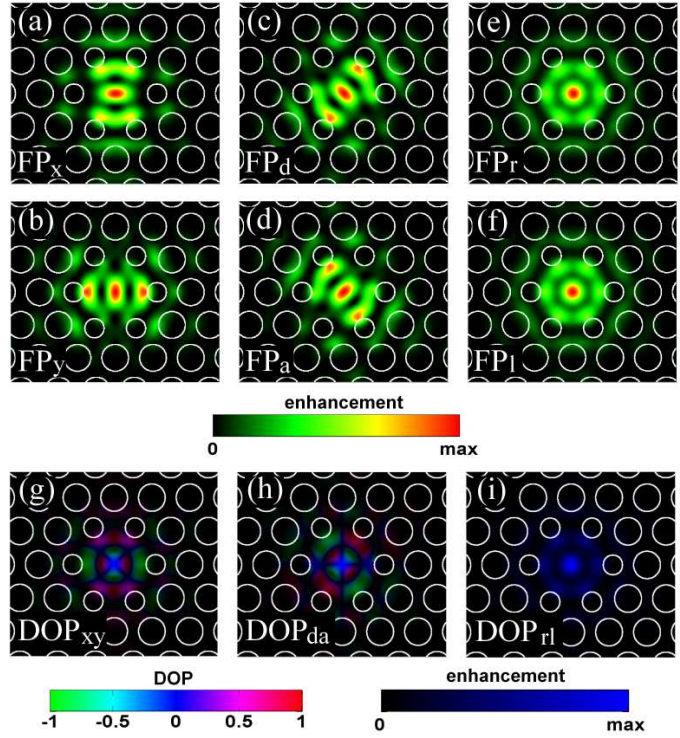}
    \caption {Spatial map of the Purcell factor for H1 cavities for (a) $Fp_x$ (b) $Fp_y$ (c) $Fp_d$ (d) $Fp_a$ (e) $Fp_r$ and (f) $Fp_l$. (g-i)
    indicate the degrees of polarization for X/Y, D/A and R/L dipoles, where the hue shows the degree of polarization and the value (intensity)
    indicates the enhancement}
    \label{fig:H1_enhancement}
\end{figure}

The DGF for the H1 cavity can be calculated using Eq.~(\ref{eq:greens_summation}) using two arbitrary orthogonal orientations of the dipole mode. From this, and using Eq.~(\ref{eq:greens_mode_func_def}) we can plot the Purcell enhancement for any chosen dipole orientation, $\bm{\mu}$, as seen in Fig.~\ref{fig:H1_enhancement}(a-f). Figure~\ref{fig:H1_enhancement}(g-i) show the DOP, as defined in Eq.~(\ref{eq:dop}), for all three bases of dipole orientations in Fig.~\ref{fig:H1_enhancement}(a-f) (rectilinear, diagonal and circular). The center of the cavity for all sources show a DOP of zero. This suggests that the response of the dipole to the cavity mode is polarization-insensitive.  This is in contrast to the L3 cavity, where the DOP is zero only for a diagonal\//anti-diagonal source in the center of the cavity.  Finally, the $DOP_{r,l} = 0$ at all points.  This is expected, as no chirality in the cavity structure is present.

\begin{figure}[htbp]
    \centering\includegraphics[width=1 \columnwidth ]{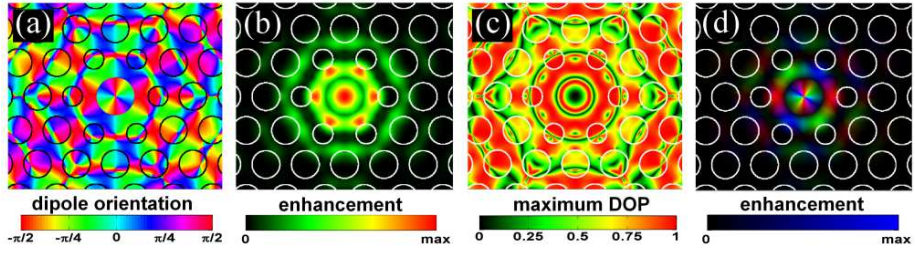}
    \caption {(a) Angle of the dipole for which maximum enhancement is achieved. (b) Enhancement of the dipole when it is oriented at the angle shown in (a), (c) maximal degree of polarization and (d) a combination of (a) and (b) where the intensity of the plot is scaled by the enhancement.}
    \label{fig:H1_max_enhancement}
\end{figure}

As with the L3 cavity, we may also calculate the optimal dipole orientation for maximum Purcell enhancement using Eq.~(\ref{eq:maximum_enhancement_angle}).  Figure~\ref{fig:H1_max_enhancement}(a) shows the optimal angle for the dipole orientation.  Using the preferred angles we can now continue to plot the maximal enhancement possible in the H1 cavity as seen in Fig.~\ref{fig:H1_max_enhancement}(b), simply by calculating the Purcell factor for the optimal dipole orientation in Fig.~\ref{fig:H1_max_enhancement}(a). Here, we note that the $C_{6v}$ symmetry is preserved. Furthermore, the only position with both a degenerate enhancement response \emph{and} a significant Purcell enhancement is in the center. The ring singularity and the six singularities between the shifted central holes occur at E-field nodes.

Next we show the maximum DOP for the preferred angles from Fig.~\ref{fig:H1_max_enhancement}(a) in Fig.~\ref{fig:H1_max_enhancement}(c). In this plot, the singularities that we mentioned in Fig.~\ref{fig:H1_max_enhancement}(a), show up again in the form of a DOP of zero, meaning there is no preferred angle of emission. From this plot it is easy to determine where a QD dipole would couple strongly to one polarization only.

Finally we combine Fig.~\ref{fig:H1_max_enhancement}(a-b) in one plot (as in Fig.~\ref{fig:L3_max_enhancement}(c)) and can be seen in Fig.\ref{fig:H1_max_enhancement}(d). From this plot it becomes clear that the dipole's response to the cavity is polarization insensitive if it is positioned perfectly in the center of the cavity.

When the QD is not perfectly positioned in the center of the cavity the spin-photon interface will not have unit fidelity. The actual fidelity of the interface is highly dependent on the physical underlying principle and the read in and read out channels. In order to give some sort of length scale, we give the ratio of the minimum ($FP_{min}$) and maximum ($FP_{max}$) enhancement, by calculating the Purcell enhancement at the optimum angle and perpendicular to it, from the values in Fig.~\ref{fig:H1_max_enhancement}(a).  This is an indication of the minimum fidelity that should be obtainable. We define this minimum fidelity as:

\begin{equation}
    \mathcal{F}=-\frac{FP_{max} - FP_{min}}{FP_{max} + FP_{min}}+1
    \label{eq:minimum_fidelity}
\end{equation}

\begin{figure}[htbp]
    \centering\includegraphics[width=0.6 \columnwidth ]{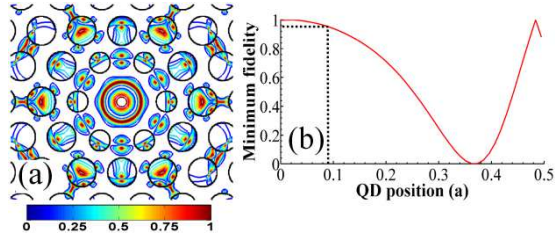}
    \caption {(a) Contour plot of the minimum fidelity of the spin-photon interface. (b) Maximum fidelity when the QD is moved along the x-axis from the center of the cavity.}
    \label{fig:H1_fidelity}
\end{figure}

Figure~\ref{fig:H1_fidelity}(a) shows a contour map of the fidelity measure from Eq.~(\ref{eq:minimum_fidelity}). A few spots can be seen where the fidelity is close to unity, however, as mentioned earlier, only the central spot shows an enhancement and is the only suitable QD location. In order to give a required precision for the QD location, we plot the fidelity against the QD position along a line in the $x$-direction as can be seen in Fig.~\ref{fig:H1_fidelity}(b). If a fidelity of 95\% is required, the QD needs to be positioned within $0.09a$ of the center. Assuming a typical resonant wavelength of 900nm, a lattice constant of $260nm$ is required for our H1 design. This means that the QD needs to be positioned within approximately 20nm of the center of the cavity.  Note that this positioning accuracy is more stringent than the $\sim 60$nm estimated to be needed for strong coupling \cite{Thom:Strong_Coupling_Positioning:APL}.  This criterion pushes the limits of what has presently been demonstrated for site controlled QDs \cite{Schneider:Site_control_rev:Nanotech}, but methods using QD identification and PhC positioning techniques have shown that the combined QD-PhC lithography positioning accuracies of $\sim 30$nm approach these requirements \cite{Thom:Strong_Coupling_Positioning:APL}.

\section{Conclusion} \label{sec:Conclusion}
In conclusion, we have demonstrated a method to determine the suitability of photonic structures for a QD spin-photon interface.  We addressed the problem from the specific point of view of QD-PhC structures for quantum information, by considering firstly in which cavity designs one is able to to create intermediate quantum superposition states, and showed that a doubly-degenerate cavity mode is required.  We then consider the influence of the QD position on the ability to excite a superposition state.  We demonstrate that even in the degenerate H1 cavity, QD position is as crucial to the superposition state formed as the dipole orientation. In particular, to allow accurate transmission of a QD spin superposition state to a cavity superposition state, the QD must be accurately placed to within $\sim 20$nm.

We note that the methods described in this paper may be applied to a quantum emitter placed in any type of lossless photonic cavity environment that modifies the LDOS, in order to calculate polarization-dependent Purcell factors for any emitter position and orientation. In particular, more complex cavity designs (such as coupled cavity, or cavity waveguide designs) may be analyzed using simple FDTD simulations and these techniques for suitability for generating and transmitting superposition states.

\section*{Acknowledgments} \label{sec:Acknowledgements}
This work was carried out using the computational facilities of the Advanced Computing Research Centre, University of Bristol -- http://www.bris.ac.uk/acrc/. This work has been funded by the project ``SPANGL4Q'', which acknowledges the financial support of the Future and Emerging Technologies (FET) programme within the Seventh Framework Programme for Research of the European Commission, under FET-Open grant number: FP7-284743.  RO is sponsored by the EPSRC as a Career Acceleration Fellow under grant no. EP/G004366/1, and JGR is sponsored under ERC Grant No. 247462 QUOWSS. We thank Mark~R.~Dennis for useful discussions.

%%%%%%%%%%%%%%%%%%%%%%% References %%%%%%%%%%%%%%%%%%%%%%%%%

\end{document}